\documentclass{moriond}

\usepackage{amsmath,amssymb}

\def\Journal#1#2#3#4{{#1} {\bf #2}, #3 (#4)}

\def\be{\begin{equation}}
\def\ee{\end{equation}}
\def\bea{\begin{eqnarray}}
\def\eea{\end{eqnarray}}

\newcommand{\Photo}{}

\begin{document}
\vspace*{4cm}
\title{Cosmology and modified gravity with dark sirens from GWTC-3}

\author{Michele Mancarella$^{1,}$\footnote{\href{michele.mancarella@unige.ch}{michele.mancarella@unige.ch} \\ To appear in the proceedings of the 56th Rencontres de Moriond 2022}, Andreas Finke$^{1}$, Stefano Foffa$^{1}$, Edwin Genoud-Prachex$^{2}$, Francesco Iacovelli$^{1}$, Michele Maggiore$^{1}$ }

\address{$^{1}$D\'epartement de Physique Th\'eorique and Center for Astroparticle Physics,\\
Universit\'e de Gen\`eve, 24 quai Ansermet, CH--1211 Gen\`eve 4, Switzerland \\
$^{2}$Institute for Theoretical Physics, Goethe University, 60438 Frankfurt am Main, Germany
}

\maketitle\abstracts{
We present the latest measurements of the Hubble parameter and of the parameter $\Xi_0$ describing modified gravitational wave propagation, obtained from the third gravitational wave transient catalog, GWTC-3, using the correlation with galaxy catalogs and information from the source-frame mass distribution of binary black holes. The latter leads to the tightest bound on $\Xi_0$ so far, i.e. $\Xi_0 = 1.2^{+0.7}_{-0.7}$ with a flat prior on $\Xi_0$, and $\Xi_0 = 1.0^{+0.4}_{-0.8}$ with a prior uniform in $\log\Xi_0$ (Max posterior and $68\%$ HDI). The measurement of $H_0$ is dominated by the single bright siren GW170817, resulting in $H_0=67^{+9}_{-6} \, \rm km \, s^{-1} \, Mpc$ when combined with the galaxy catalog.}
\section{Introduction}
Gravitational Waves (GWs) from coalescing binaries are direct distance tracers, as the luminosity distance is measured directly from the GW signal~\cite{Schutz}. Combined with redshift information, this allows at the same time to measure the expansion history of the Universe and to test General Relativity (GR) at cosmological scales. The reason is that any modification of GR at cosmological scales leads to extra friction experienced by GWs during their propagation, which results in a modification of the notion of luminosity distance as measured by GWs (known as ``modified GW propagation''). On a $\Lambda$CDM background and at late times, such distance can be parametrized as~\cite{Belgacem:2018lbp}
\begin{equation}\label{dLemmod}
d_L^{\rm GW}(z)= \Bigg[\Xi_0 +\frac{1-\Xi_0}{(1+z)^n} \Bigg]\times \frac{c}{H_0}\, (1+z) \,\int_0^z\, 
\frac{d\tilde{z}}{\sqrt{\Omega_{\rm m,0} (1+\tilde{z})^3+1-\Omega_{\rm m,0}}}\, .
\end{equation}
The parameters $(\Xi_0, n)$ encode the effects of non-standard friction with respect to GR (defined by $\Xi_0=1$).
The redshift of the source, whose knowledge is crucial to test the distance-redshift relation~(\ref{dLemmod}), cannot be determined by the GW signal alone due to a well-known degeneracy between source-frame mass and redshift in the GW waveform. 
In absence of a direct electromagnetic (EM) counterpart to the GW event, the source goes under the name of ``dark siren'' and statistical techniques have to be adopted to obtain the redshift information.

In this contribution, we present the application of two such techhiques to the latest gravitational wave transient catalog GWTC-3~\cite{GWTC3} with new, fully independent, open-source codes: the correlation with galaxy catalogs and the use of information from the source-frame mass distribution of binary black holes.

\section{Population studies and dark sirens}\label{sec:methods}

Statistical information on the redshift can be added within a hierarchical Bayesian analysis of the population~\cite{Mandel:2018mve}.
The population of GW sources is described by a population function $p_{\rm pop}(\theta | \Lambda)$ giving the probability that a source has parameters $\theta$ given hyperparameters $\Lambda$. 
This is known as a distribution in redshift and source-frame masses~\footnote{In this work we neglect the spins that can be treated in an analogous way.}, parametrized by a set of parameters $\Lambda_{\rm astro}$, while GW experiments measure detector-frame quantities. The conversion between the two relies on the distance-redshift relation~(\ref{dLemmod}), hence $p_{\rm pop}$ acquires a dependence on the parameters of this relation, that we denote by $\Lambda_{\rm cosmo}$. Thus $\Lambda = \{\Lambda_{\rm astro}, \Lambda_{\rm cosmo}\}$. 
One can write a likelihood for the parameters $\Lambda$ by marginalising over the single GW event likelihood $p(\mathcal{D}_i| \theta_i)$, re-weighted by the population prior, as follows~\cite{Mandel:2018mve}:
\begin{eqnarray}\label{likelihood}
  p(\mathcal{D} | \Lambda ) \propto 
  \prod_{i=1}^{N_{\rm obs}} \frac{1}{\alpha(\Lambda)} \int d\theta_i \,  p(\mathcal{D}_i| \theta_i) \, p_{\rm pop}(\theta_i | \Lambda)  \, .
 \end{eqnarray}
In the above equation, the term $\alpha(\Lambda)$ is the the fraction of expected detections and corrects for selection bias, i.e. the fact that the likelihood for a GW experiment to observe an event varies strongly depending on the source parameters. Not accounting for this would result in a biased measurement when analyzing a population, and a correct modeling of the latter has to be included in any analysis~\cite{Mandel:2018mve}. 
The interplay between detector- and source-frame quantities in the likelihood~(\ref{likelihood}) allows breaking the mass-redshift degeneracy and constrain the parameters $\Lambda_{\rm cosmo}$. 
Given the limited statistical power of current data, in this contribution we consider separately two cases: (i) the expansion history is inferred within $\Lambda$CDM, i.e. $\Xi_0$ is fixed to 1 and $ \Lambda_{\rm cosmo} = \{H_0, \Omega_{\rm m,0}\}$; and (ii) the expansion history is fixed by Planck 2018, and $ \Lambda_{\rm cosmo} = \{\Xi_0, n\}$.
\section{Results from the binary black hole mass distribution}\label{sec:BBHmass}
\begin{figure}[t]
\centering
\includegraphics[height=5cm,keepaspectratio]{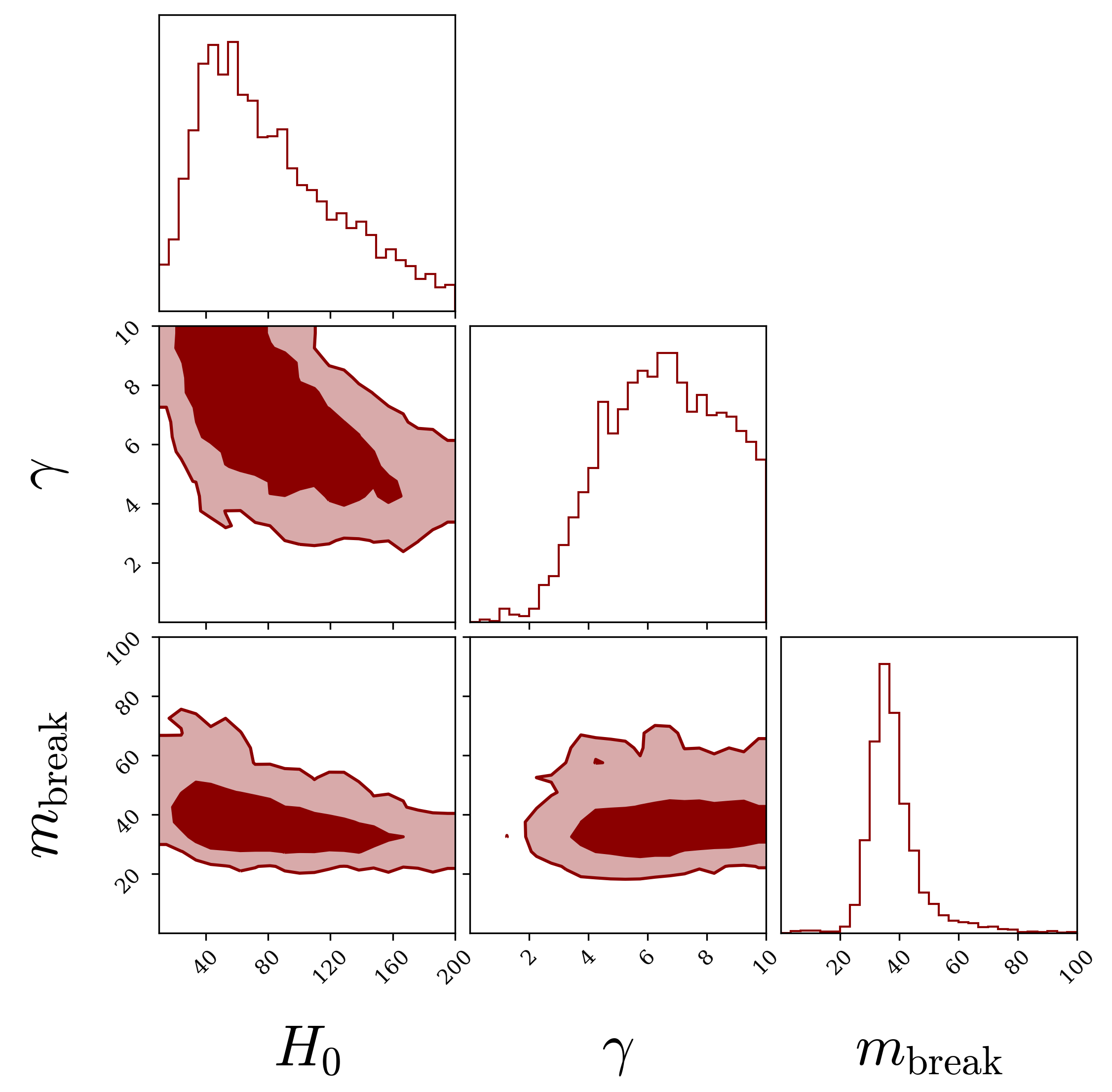}
\hspace{2cm}
\includegraphics[height=5cm,keepaspectratio]{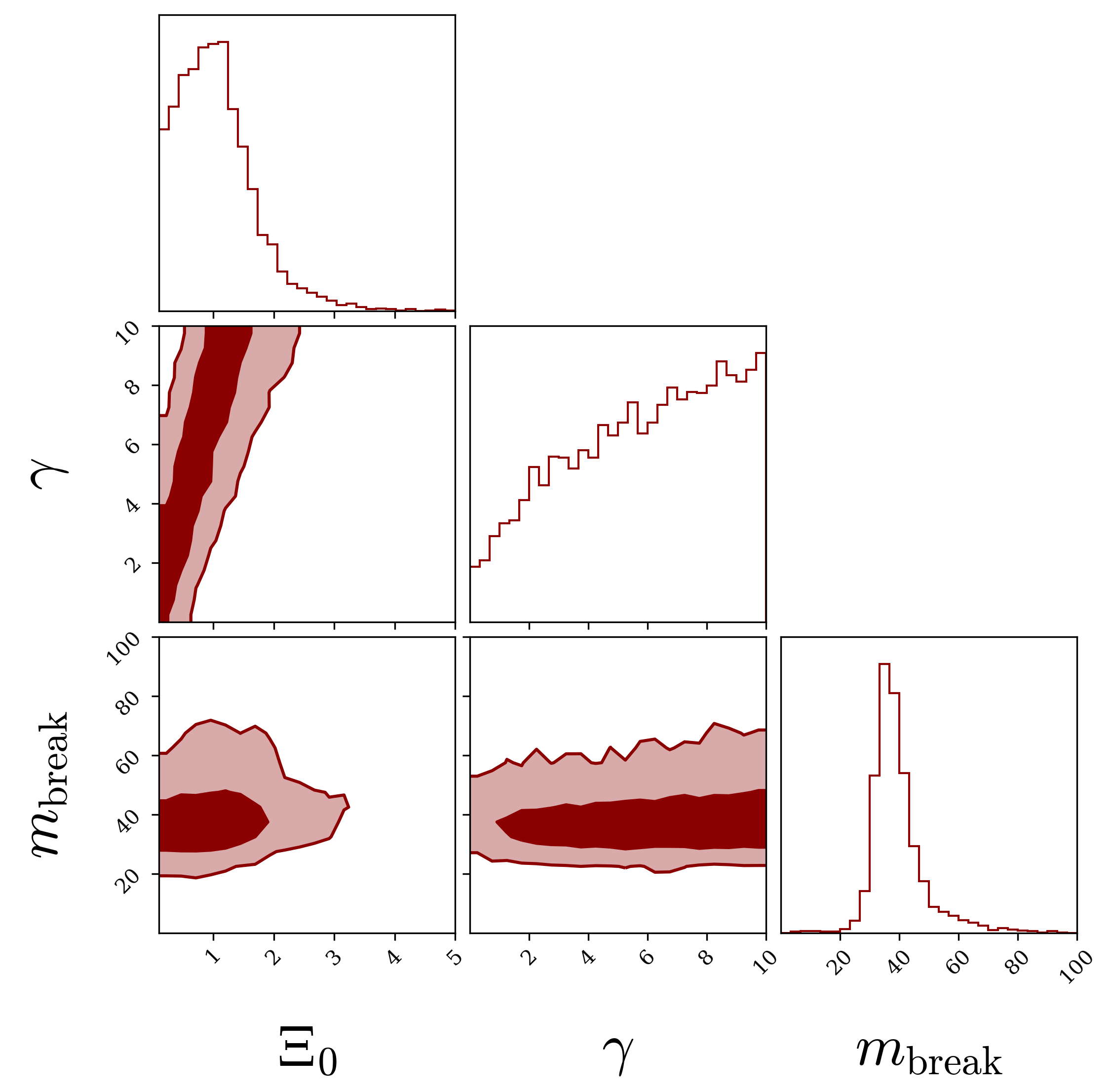}
\caption{Joint constraints on population and cosmological parameters assuming a feature in the BBH mass distribution at a scale $m_{\rm break}$ and a BBH merger rate evolution $\sim (1+z)^{\gamma}$ at low $z$. All the parameters not appearing in the corner plots have been marginalised over. Left: constraint on $H_0$ within $\Lambda$CDM ($\Xi_0=1$). Right: constraint on $\Xi_0$, fixing the expansion history to Planck 2018.}
\label{fig:massFunction}
\end{figure}
The mass distribution of stellar origin Binary Black Holes (BBHs) features a drop-off between $\sim40-60  M_{\odot}$, which reflects the imprint of the pair-instability supernovae process, according to which BH remnants above this scale are not produced if the BH progenitor is a Helium star in the $\sim40-120 M_{\odot}$ range~\cite{SperaMapelli19}. This scale can be included in $p_{\rm pop}$ to break the mass-redshift degeneracy~\cite{Farr:2019twy}.
We adopt a parametric form for the source-frame mass distribution given by the ``broken power law model'', where the mass scale is encoded in a parameter $m_{\rm break} \in \Lambda_{\rm astro} $.
The parameters of the source frame distribution are not known \emph{a priori} and are subject to large uncertainties, so the sets $\{\Lambda_{\rm astro}, \Lambda_{\rm cosmo}\}$ have to be inferred simultaneously.
We apply this technique to GWTC-3 using a sample of 35 events with network signal-to-noise-ratio (SNR) larger than $12$. We assume a parametric redshift distribution that follows the Madau-Dickinson rate, scaling as $\sim (1+z)^{\gamma}$ at low redshift. We refer to Ref.~\cite{Mancarella:2021ecn} for all details of the analyis, including methodology, precise definitions, prior choices, and computation of the selection bias. Fig.~\ref{fig:massFunction} shows the constraints for case (i) (left) and case (ii) (right).~\footnote{The corner plot is restricted to the subset of the parameter space with physically relevant correlations to the cosmological parameters, i.e. the scale $m_{\rm break}$ and the rate evolution parameter $\gamma$. See Ref.~\cite{Mancarella:2021ecn} for the complete results.} 
A mass scale is detected around $\sim 30-45 M_{\odot}$, which drives the constraint on the cosmological parameters. The rate evolution remains instead much less constrained, in particular for case (ii), which is explained by the fact that the effect of $\Xi_0$ evolves with redshift and is thus more degenerate with the effect of $\gamma$.~\footnote{See Ref.~\cite{Mancarella:2021ecn} for a detailed study of this effect and the impact of $\Xi_0$ on the reconstructed BBH merger rate.} We obtain $\Xi_0 = 1.2^{+0.7}_{-0.7}$ with a flat prior on $\Xi_0$, while using a prior uniform in $\log\Xi_0$ we find $\Xi_0 = 1.0^{+0.4}_{-0.8}$ (shown in Fig.~\ref{fig:massFunction}). For the Hubble paramter, we find $H_0=50^{+53}_{-26} \, \rm km \, s^{-1} \, Mpc$.~\footnote{All uncertainties quoted are max posterior and $68\%$ HDI.}
\section{Results from the correlation with a galaxy catalog}
\begin{figure}[t]
\centering
\includegraphics[height=4cm,keepaspectratio]{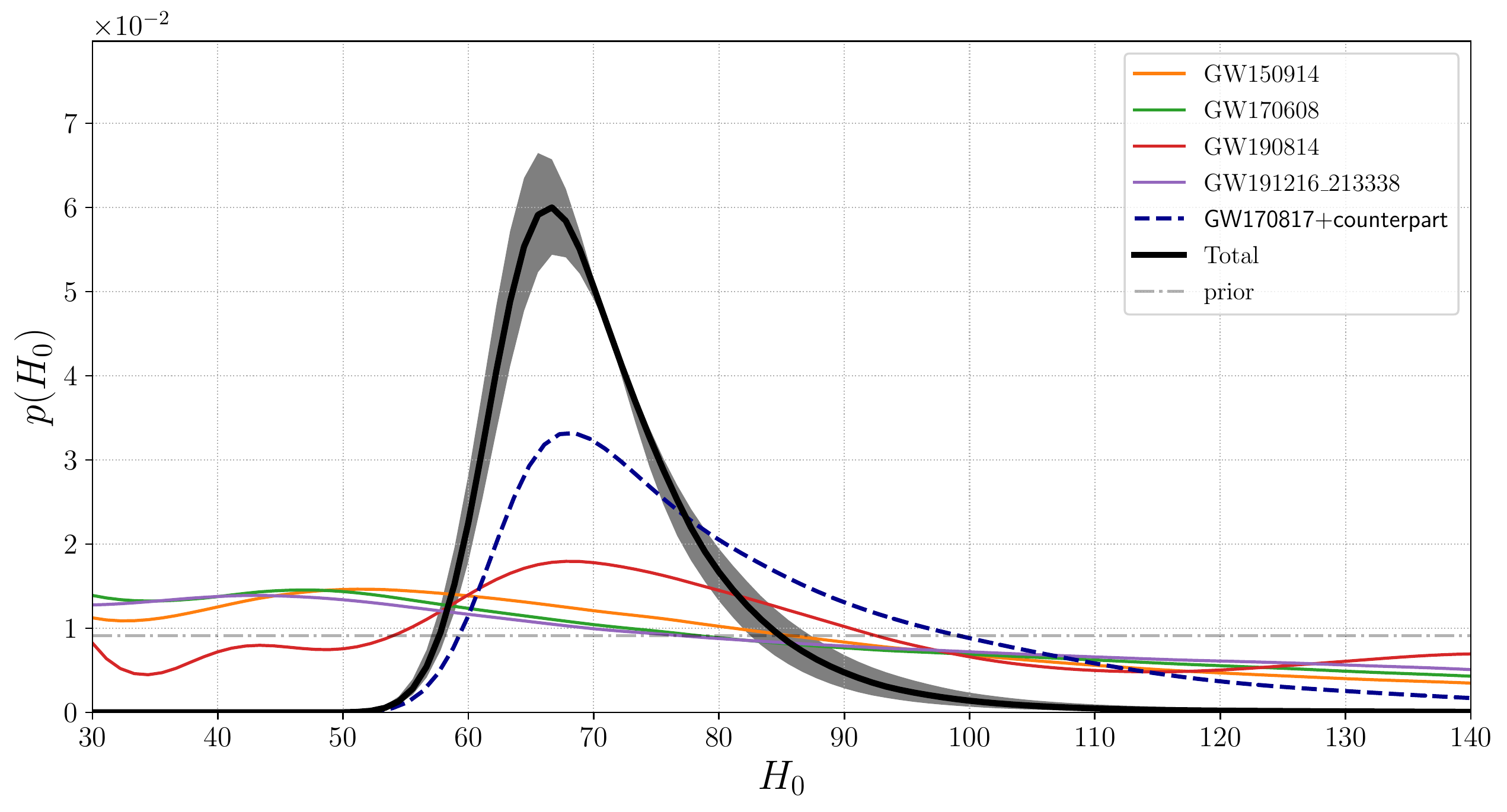}
\includegraphics[height=4cm,keepaspectratio]{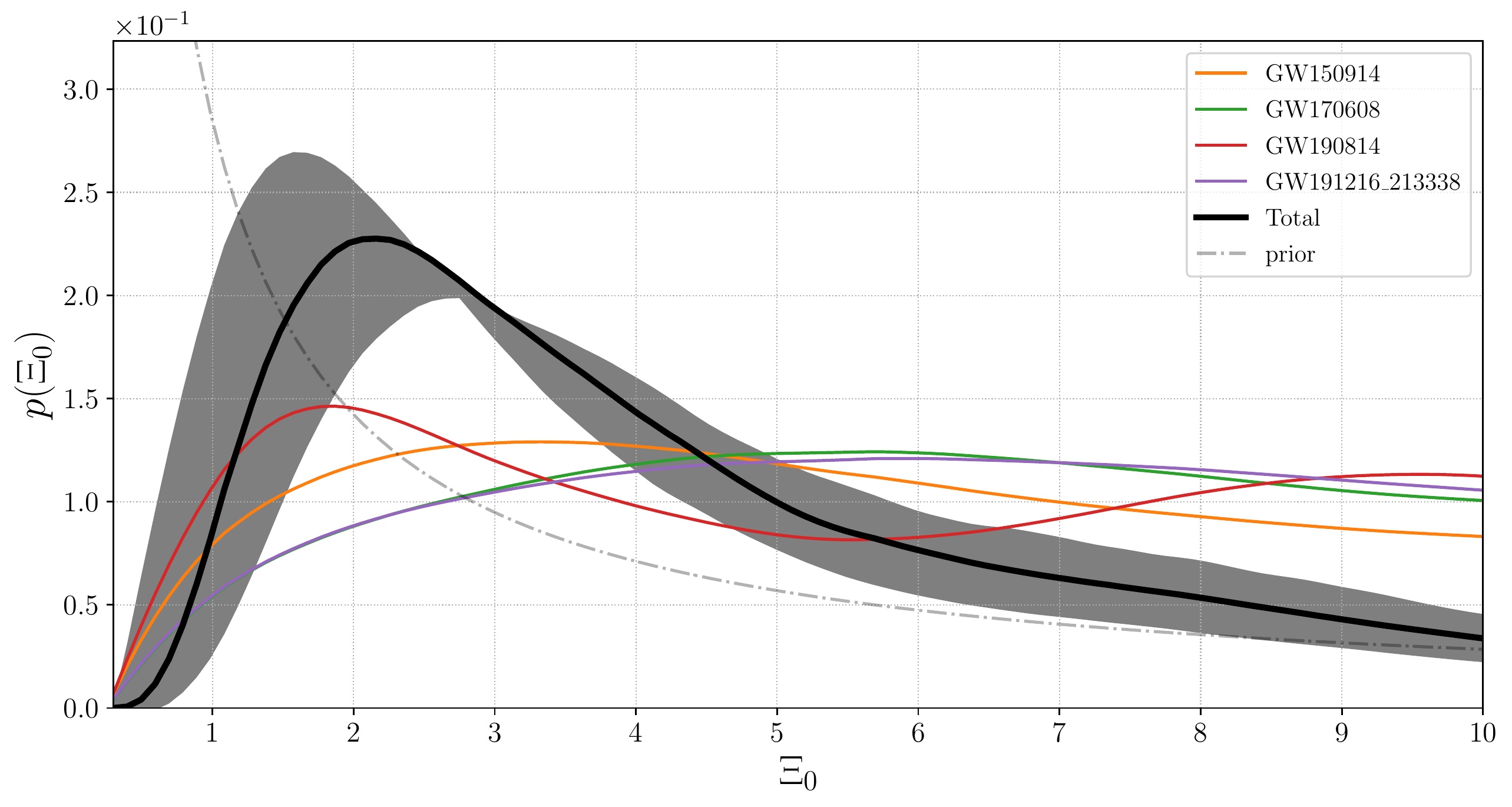}
\caption{Constraints from the galaxy catalog method and, for $H_0$, from the detection of the EM counterpart to GW170817. The colored lines are the posterior from individual events, while the black solid line the combined constraint. The dark band shows the variation of the posterior if we vary the parameter $\gamma$ of the Madau-Dickinson rate within the allowed range obtained from the analysis of Sec.~\ref{sec:BBHmass} (see the text for a discussion).
Left: constraint on $H_0$ within $\Lambda$CDM ($\Xi_0=1$). Right: constraint on $\Xi_0$, fixing the expansion history to Planck 2018.}
\label{fig:galcat}
\end{figure}
A way to obtain prior knowledge on the redshift is to include in the population function a redshift distribution computed from a catalog of all the galaxies in the GW localization volume, and marginalise over the choice of the galaxy~\cite{Schutz}. 
In principle, a joint inference including also the BBH mass distribution as in Sec.~\ref{sec:BBHmass} would be necessary, but the computational cost of such analysis is currently too high. 
Hence, here we shall consider two special cases of case (i) and (ii) above, where we further fix all parameters except either $H_0$ or $\Xi_0$.
If the catalog is not complete, either the prior distribution on redshift is supplemented by a suitable ``completion'' accounting for the missing galaxies, or the analysis is restricted to events falling in complete regions. In the latter case, however, this introduces an additional selection effect that must be accounted for in the computation of the selection bias~\cite{{Finke:2021aom}}. 
The open-source code $\tt{DarkSirensStat}$~\cite{{Finke:2021aom}} allows both possibilities. We refer to Ref.~\cite{{Finke:2021aom}} for a thorough description of the methodology applied here and implemented in the code.
We analyse the GWTC-3 catalog using this code and the ``GLADE+'' galaxy catalog~\cite{Dalya:2021ewn}, with the same SNR threshold of Sec.~\ref{sec:BBHmass} and restricting to events whose position is in a region where the catalog is $100\%$ complete~\footnote{Other settings for the analyisis, except those explicitly indicated here, are taken equal to the default settings of the results in Ref.~\cite{Finke:2021aom} }.
In this case, we also include in the analysis the event GW190814, which has a low secondary mass and for this reason was excluded in the analysis of Sec.~\ref{sec:BBHmass}, but is currently the best localised dark siren.
To obtain the population parameters, we first analyse GWTC-3 with the technique of Sec.~\ref{sec:BBHmass} using the open-source code $\tt{MGCosmoPop}$~\cite{Mancarella:2021ecn}, including this time the event GW190814~\footnote{The prior choices are the same as in Ref.~\cite{Mancarella:2021ecn}, with a prior uniform in $\log \Xi_0$, except for the following choices adopted here:$\gamma \in [0, 10]$, $\Xi_0 \in [0.3, 10]$, $H_0 \in [30, 140]  \, \rm km \, s^{-1} \, Mpc $ .}. Then, we run $\tt{DarkSirensStat}$ fixing the population parameters to those obtained with $\tt{MGCosmoPop}$~\footnote{We modified the code $\tt{DarkSirensStat}$ to include the Madau-Dickinson rate and to match the SNR criterion adopted here in the computation of selection effects. }.
Fig.~\ref{fig:galcat} shows the results for $H_0$ (left) and for $\Xi_0$ (right) fixing all the other parameters. 
In the case of $H_0$, we also combine with the result obtained from the detection of the Binary Neutron Star GW170817 and its counterpart GRB 170817A~\cite{LIGOScientific:2017vwq}, obtained as detailed in Sec. 4.1.2 of Ref.~\cite{Finke:2021aom}.
To estimate the effect of population uncertainty, we repeat the analysis varying the parameter $\gamma$ describing the BBH merger rate evolution with redshift (which is the one with the largest impact) within the $68\%$ C.L. obtained from the population analysis. For case (i) ($H_0$ with fixed $\Xi_0$), this is  $\gamma=7.0^{+1.9}_{-1.9}$. For case (ii) ($\Xi_0$ with fixed $H_0$), we have $\gamma=5.7^{+2.9}_{-3.3}$. Effects of the variation of $\gamma$ are shown in the gray band in Fig.~\ref{fig:galcat}. The result for dark sirens is largely prior-dominated, mostly due to the large GW localization regions and to the incompleteness of the catalog, which limits the useful events to only four. For the fiducial values of the population, we obtain 
$\Xi_0=2.2^{+2.9}_{-1.1}$ and $H_0=67^{+9}_{-6} \, \rm km \, s^{-1} \, Mpc$ (Max posterior and $68\%$ HDI).\footnote{For $\Xi_0$, the $90\%$ HDI is $\Xi_0=2.2^{+5.8}_{-1.3}$. } 
\section{Summary and outlook}
We presented state-of-the-art constraints on the Hubble parameter and on the parameter $\Xi_0$ describing modified GW propagation, obtained with dark siren techniques from the GWTC-3 catalog. The presence of a feature in the BBH mass function gives the tightest bound on $\Xi_0$, $\Xi_0 = 1.2^{+0.7}_{-0.7}$ with a flat prior on $\Xi_0$, and $\Xi_0 = 1.0^{+0.4}_{-0.8}$ with a prior uniform in $\log\Xi_0$.
The main systematics of this method, to be addressed as the statistical uncertainty reduces, are the correct modeling of the population (including in particular the possible evolution of the mass function with redshift) and the presence of outliers. 
The tightest measurement of $H_0$ is driven by the single bright siren GW170817, combined with correlation with the galaxy catalog, which gives $H_0=67^{+9}_{-6} \, \rm km \, s^{-1} \, Mpc$ (Max posterior and $68\%$ HDI).
In general, the correlation with ``GLADE+'' alone is much less constraining due to the large localization volumes of GW events, incompleteness of the catalog, and uncertainty in the population model. The latter is treated so far as a source of systematic uncertainty, but this should be turned in a statistical uncertainty within a joint astrophysical and cosmological analysis, which however poses non trivial computational challenges. Addressing these challenges, using more data from upcoming observing runs of the GW observatories, using more complete catalogs, and exploring other statistical techniques and their combinations, are extraordinary avenues for GW cosmology that can lead to substantial advances in the coming years.


\section*{References}

\begin{thebibliography}{99}

\bibitem{Schutz} 
B. F. Schutz, \Journal{\em Nature}{323}{}{1986}, doi:\href{https://doi.org/10.1038/323310a0}{10.1038/323310a0}

 \bibitem{Belgacem:2018lbp} 
E. Belgacem,  Y. Dirian, S. Foffa, M. Maggiore, \Journal{\em Phys. Rev. D}{98}{023510}{2018}, doi:\href{https://doi.org/10.1103/PhysRevD.98.023510}{10.1103/PhysRevD.98.023510}, \href{https://arxiv.org/abs/1805.08731}{arXiv:1805.08731}

 \bibitem{GWTC3}
 LIGO Scientific, VIRGO, and KAGRA collaborations, (2021), \href{https://arxiv.org/abs/2111.03606}{arXiv:2111.03606}

\bibitem{Mandel:2018mve} 
I. Mandel, W. M. Farr, J. R. Gair, \Journal{\em Mon. Not. Roy. Astron. Soc.}{486}{1}{2019}, doi:\href{https://doi.org/10.1093/mnras/stz896}{10.1093/mnras/stz896}, \href{https://arxiv.org/abs/1809.02063}{arXiv:1809.02063}


\bibitem{Farr:2019twy} 
W. Farr, M. Fishbach, J. Ye, D. Holz, \Journal{\em Astrophys. J. Lett.}{883}{L42}{2019}, doi:\href{https://doi.org/10.3847/2041-8213/ab4284}{10.3847/2041-8213/ab4284}, \href{https://arxiv.org/abs/1908.09084}{arXiv:1908.09084}


\bibitem{SperaMapelli19}
M. Spera, M. Mapelli, \Journal{\em Mon. Not. Roy. Astron. Soc.}{470}{4}{2017}, doi:\href{https://doi.org/10.1093/mnras/stx1576}{10.1093/mnras/stx1576}, \href{https://arxiv.org/abs/1706.06109}{arXiv:1706.06109}

\bibitem{Mancarella:2021ecn}M. Mancarella, E. Genoud-Prachex, M. Maggiore, \Journal{\em Phys. Rev. D}{}{To appear}{2021},  \href{https://arxiv.org/abs/2112.05728}{arXiv:2112.05728}

\bibitem{Finke:2021aom}
 A. Finke, S. Foffa, F. Iacovelli, M. Maggiore, and M. Mancarella, \Journal{\em JCAP}{2108}{026}{2021}, doi:\href{https://iopscience.iop.org/article/10.1088/1475-7516/2021/08/026}{10.1088/1475-7516/2021/08/026}, \href{https://arxiv.org/abs/2101.12660}{arXiv:2101.12660}

\bibitem{LIGOScientific:2017vwq}
LIGO Scientific and VIRGO collaborations, \Journal{\em Phys. Rev. Lett.}{119}{161101}{2017}, doi:\href{https://doi.org/10.1103/PhysRevLett.119.161101}{10.1103/PhysRevLett.119.161101}, \href{https://arxiv.org/abs/1710.05832}{arXiv:1710.05832}

\bibitem{Dalya:2021ewn}
G. D\'alya \emph{et al.} (2021), \href{https://arxiv.org/abs/2110.06184}{arXiv:2110.06184}


\end{thebibliography}
\end{document}